\documentclass[groupedaddress,twocolumn]{revtex4-2} 

\usepackage[utf8]{inputenc}
\usepackage[T1]{fontenc}

\usepackage{amsmath}
\usepackage{amssymb}
\usepackage{amsfonts}
\usepackage{amsthm}
\usepackage{bm}
\usepackage{braket}
\usepackage{physics}
\usepackage{mathtools}
\usepackage[normalem]{ulem}
\usepackage{graphicx}
\usepackage{dcolumn}
\usepackage{xcolor}
\usepackage{float}
\usepackage{subcaption}
\usepackage{comment}

\usepackage[colorlinks=true,
linkcolor=blue,
citecolor=blue,
urlcolor=blue]{hyperref}

\usepackage[colorinlistoftodos]{todonotes}
\usepackage[most]{tcolorbox}

\newtcolorbox{comentario}{
    colback=yellow!10,
    colframe=gray!70!black,
    title=\textbf{Comentario},
    fonttitle=\bfseries,
    boxrule=0.6pt,
    arc=2mm,
    left=2mm,
    right=2mm,
    top=1mm,
    bottom=1mm
}

\begin{document}

\title{
Robustness of Dynamical Casimir Effect-Induced Quantum Synchronization}

\author{Sebastián Francisquez}
\email{s.francisquez@df.uba.ar}

\author{Fernando C. Lombardo}
\author{Paula I. Villar}
\affiliation{
Departamento de Física Juan José Giambiagi, FCEyN UBA and IFIBA UBA-CONICET,
Facultad de Ciencias Exactas y Naturales, Ciudad Universitaria, Pabellón I
}
\date{\today}


\begin{abstract}
The robustness of quantum synchronization induced by the dynamical Casimir effect (DCE) is analyzed within a circuit quantum electrodynamics (cQED) architecture consisting of two superconducting qubits coupled to a shared, parametrically driven cavity. Using a Lindblad master equation approach, this study evaluates the impact of various decoherence channels (including photon loss, relaxation, thermal excitation, and pure dephasing) on the resulting synchronization dynamics. The results demonstrate that DCE-induced synchronization persists under dissipation rates achievable in state-of-the-art superconducting platforms, confirming its experimental feasibility. Among the studied mechanisms, pure dephasing is identified as the dominant factor limiting the fidelity of the synchronized state. Furthermore, it is shown that in strongly dissipative regimes, the Pearson correlation coefficient may overestimate the degree of quantum synchronization, as environmental relaxation can generate classical correlations that mimic synchronized behavior. These findings establish the parameter regimes necessary for observing genuine DCE-induced quantum synchronization and provide practical guidelines for future cQED experiments

\end{abstract}

\maketitle


\section{Introduction}
Since Huygens first observed the synchronization of
pendulum clocks \cite{hyugens1,hyugens2},
synchronization has been
recognized
in a wide variety of systems. Examples of this phenomenon include
biological systems, such as the
synchronous flashing of fireflies
\cite{luciernagas}, and
neural
systems, where synchronization plays a central role in information
processing \cite{neurociencia}. The
occurrence
of synchronization in systems of such
diverse nature
highlights
its universal character.

In recent years, this concept has been extended to the quantum regime
\cite{sincro1,sincro2,sincro3,sincro4,sincro5}, where the dynamics is influenced by genuinely quantum features such as quantum fluctuations, coherence, and entanglement, making
the characterization of quantum synchronization
nontrivial. As a consequence, quantum synchronization exhibits features
with no classical analogue and requires specific conceptual tools for
its description \cite{Solanki2026}.

While quantum coherence and vacuum fluctuations are intrinsic features of quantum systems, quantum correlations such as entanglement are generated only under specific conditions. The relationship between these quantum resources and synchronization remains under debate, since some works report that quantum correlations favor synchronization \cite{mejora_sinc}, while others show that they may hinder its emergence \cite{inhibicion}, suggesting that there is no simple or universal relationship between them. Nevertheless, studying this relationship is of both conceptual and practical interest. Quantum synchronization has potential applications in communication, precise timekeeping, advanced navigation, sensing, and quantum information processing, as well as in the implementation of integrated real-time synchronization devices \cite{aplicacion2,aplicacion3,aplicacion4,aplicacion5}.

Among the mechanisms capable of generating quantum correlations, the dynamical Casimir effect (DCE) has emerged as a promising resource. The DCE originates from vacuum fluctuations of quantum fields, revealing that the quantum vacuum possesses observable physical consequences \cite{casimirestatico}. When the boundary conditions of a field vary in time, these vacuum fluctuations can be amplified, giving rise to real excitations through the DCE \cite{DCE}. In the electromagnetic case, this phenomenon can be interpreted as the generation of photons from the vacuum through a nonadiabatic modulation of the field boundary conditions.

Superconducting circuits provide a natural platform for the experimental
implementation of the DCE. In \cite{medicionDCE}, effects associated with
the DCE
were observed in a superconducting coplanar waveguide terminated by a
SQUID. Superconducting qubits operate
at very low temperatures, on the order of 10--30 mK, where the thermal
population of the resonator is extremely low, minimizing thermal noise
effects
\cite{fotonestermicos1,fotonestermicos2,fotonestermicos3,fotonestermicos4,fotonestermicos5,fotonestermicos6}.
In this regime,
quantum effects such as the DCE become experimentally accessible, making the DCE an exploitable quantum resource.

In Ref.~\cite{generatephotons}, it was shown that even in configurations involving two atoms inside a cavity, the DCE can generate photons and increase the probability of atomic excitation. Subsequently, several works demonstrated that DCE can also be used to generate entanglement~\cite{entanglement1,entanglement2,entanglement3,entanglement4}. In Ref.~\cite{tesisjapon}, the implementation of DCE to induce quantum synchronization between two qubits was studied, establishing the DCE as a viable mechanism for inducing synchronization between qubits. Despite these advances, the robustness of DCE-induced synchronization against losses and decoherence has not yet been systematically studied, leaving open a fundamental question regarding its viability under realistic dissipative conditions.

In this work, we extend the analysis of Ref.~\cite{tesisjapon} to the framework of open quantum systems. We investigate the robustness of DCE-induced synchronization under realistic environmental conditions relevant to circuit quantum electrodynamics by analyzing cavity losses, qubit relaxation, and pure dephasing. In particular, we determine the range of experimentally accessible decay rates for which synchronization remains stable.


However, as environmental losses increase, dissipative processes can generate correlations that mimic synchronized behavior, making synchronization measures alone insufficient to fully characterize the system dynamics. In this regime, additional quantifiers of quantum correlations, such as quantum discord, become necessary to distinguish genuine synchronization from correlations induced by the environment. Furthermore, we identify dephasing as the dominant decoherence mechanism limiting synchronization, providing information on the conditions required for the experimental observation of DCE-induced synchronization.

This article is organized as follows: In the next section, we present the model under study, where two decoupled qubits are coupled to a single-mode cavity implemented in a circuit quantum electrodynamics (cQED) platform based on superconducting circuits. In the same section, we show how this system should be treated when dissipation and decoherence are taken into account through the use of master equations in the Lindblad form. In Section III, we present and discuss the main measures used to establish the degree of quantum synchronization between the qubits, while in Section IV, we present the main results of our work, analyzing the robustness conditions of quantum synchronization in cQED implementations and studying the role of the different dissipation and dephasing channels in the model. Section V contains our main conclusions.

\section{Model}

\subsection{Closed-system Hamiltonian}

In this section, we present the model used to describe the dynamics of two qubits coupled to a single-mode cavity. In the following, we shall  adopt the convention $\hbar=1$.

\begin{figure*}[t]
\centering
\includegraphics[width=0.95\textwidth]{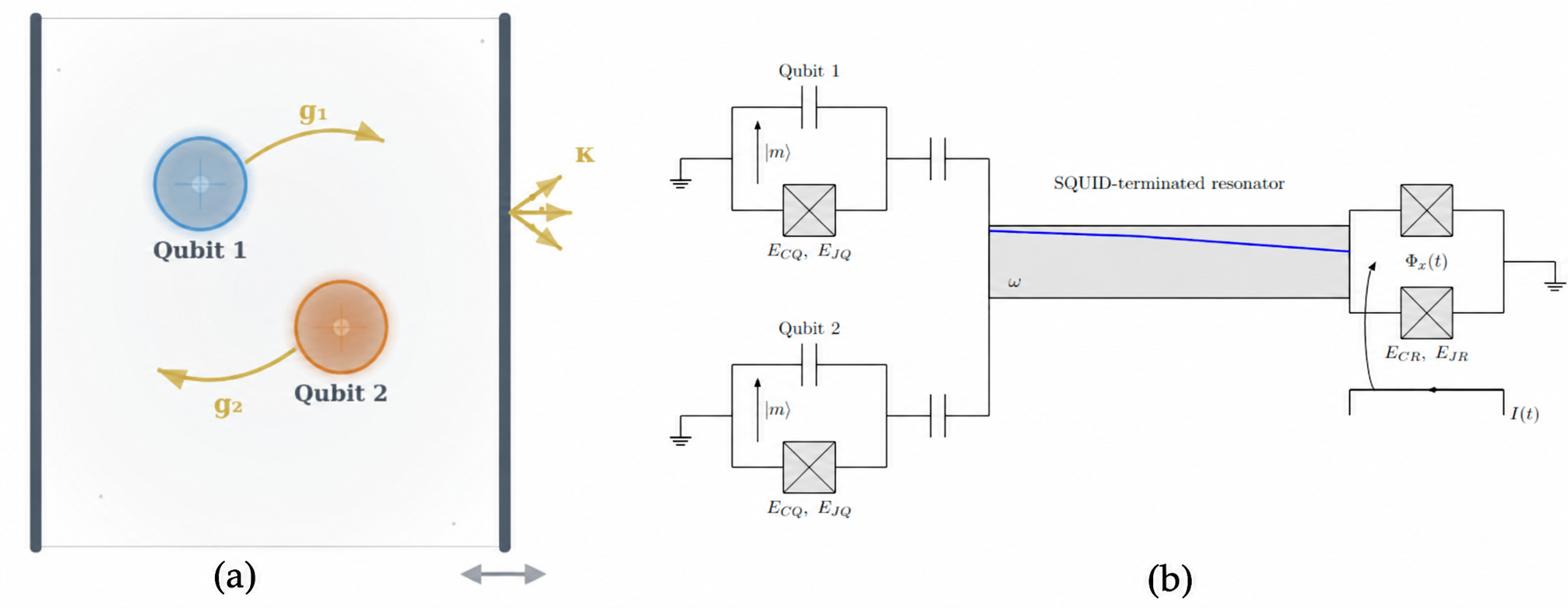}
\caption{General scheme of the setup. (a) Schematic representation of the unitary model. The system consists of two qubits coupled to a single cavity mode. The qubits do not interact directly with each other, and their only interaction is mediated by the cavity through the coupling strengths $g_1$ and $g_2$. The mirror is being moved to generate photons via the dynamical Casimir effect.(b) Circuit implementation of the model. Two transmon qubits are coupled to a transmission line terminated by a SQUID. The time-dependent external magnetic flux applied to the SQUID generates a time-dependent boundary condition, enabling the dynamical Casimir effect.}
\label{esquema}
\end{figure*}

The coherent dynamics of the system is governed by the Hamiltonian

\begin{equation}
\hat{H}=\hat{H}_{\rm TC}+\hat{H}_{\rm DCE},
\end{equation}
where $\hat{H}_{\rm TC}$ is the two-qubit Tavis--Cummings Hamiltonian \cite{taviscummings}, given by

\begin{equation}
\hat{H}_{\rm TC}= \omega \hat{a}^\dagger \hat{a}
+
\sum_{\mu=1}^2
\left[
\frac{1}{2}\omega_q^\mu\hat{\sigma}_z^\mu
+
g_\mu
\left(
\hat{\sigma}_-^\mu \hat{a}^\dagger
+
\hat{\sigma}_+^\mu \hat{a}
\right)
\right]
\label{taviscummings}
\end{equation}
and $\hat{H}_{\rm DCE}$ describes the dynamical Casimir effect (DCE) \cite{HDCE} given by
\begin{equation}
\hat{H}_{\rm DCE}=\alpha(t)(\hat{a}^\dagger+\hat{a})^2.
\label{HDCE}
\end{equation}

In the equations above, $\hat a$ ($\hat a^\dagger$) denotes the annihilation (creation) operator of the cavity mode with frequency $\omega$, while $\hat{\sigma}_z^\mu$ and $\hat{\sigma}_{\pm}^\mu$ are the Pauli and raising/lowering operators of qubit $\mu$, whose transition frequency is $\omega_q^\mu$. The parameters $g_\mu$ describe the coupling strength between each qubit and the cavity.

A schematic representation of the system is shown in Fig.~\ref{esquema}, together with a physical implementation based on superconducting circuits. The electromagnetic field is confined in a coplanar-waveguide resonator, which acts as a $\lambda/4$ cavity supporting discrete modes. Restricting the description to a single resonant mode, the resonator is modeled as a quantum harmonic oscillator, giving rise to the first term of Eq.~\eqref{taviscummings} \cite{medicionDCE,cavidad}. The qubits are realized by superconducting transmons consisting of a Josephson junction shunted by a large capacitance. In the weak-excitation regime, only the two lowest energy levels are relevant, allowing each transmon to be approximated as a two-level system and giving rise to the second term of Eq.~\eqref{taviscummings} \cite{fotonestermicos5}. The capacitive coupling between the transmons and the resonator results in the interaction term of the Tavis-Cummings Hamiltonian.

One end of the resonator is terminated by a superconducting quantum interference device (SQUID) whose effective inductance can be tuned through an externally applied magnetic flux. Such SQUID-terminated resonators have been extensively employed to study the dynamical Casimir effect \cite{medicionDCE,squid,squid2}. By modulating the external flux in time, the effective electrical length of the resonator becomes time dependent, resulting in a dynamical boundary condition for the cavity field. As shown in Ref.~\cite{nicoferpaz}, this time-dependent boundary condition leads to an effective parametric driving term in the cavity Hamiltonian, which can be written as Eq.~\eqref{HDCE}.

The time-dependent modulation of the boundary conditions induced by the SQUID is modeled as

\begin{equation}
\alpha(t)
= \alpha_0
\cos(\omega_d t)
\Theta(\tau-t),
\end{equation}
where $\omega_d$ is the driving frequency, $\alpha_0$ is the modulation amplitude, $\Theta(t)$ is the Heaviside step function, and $\tau$ is the total time the modulation lasts. The Heaviside function is used to turn off the modulation after a finite time $\tau$. 

To simplify the dynamics, we move to a rotating frame (equivalently, to the interaction picture with respect to $H_0$) defined by the unitary transformation $
\hat R=e^{i\hat H_0 t}
$,
with

\begin{equation}
\hat H_0=
\frac{\omega_d}{2}\hat a^\dagger \hat a
+
\sum_{\mu=1}^{2}
\frac{\omega_d}{2}
\frac{\hat\sigma_z^\mu}{2}.
\end{equation}

This transformation removes the fast oscillating dynamics associated with the driving frequency and provides a convenient starting point for the rotating-wave approximation. The transformed Hamiltonian takes the form

\begin{equation}
\begin{split}
\hat{\mathrm{H}}
=&\,
\Delta\hat N
+
\sum_{\mu=1}^2
\Bigg[
\Delta^\mu \frac{\hat{\sigma}_z^\mu}{2}
+
g_\mu
\left(
\hat{\sigma}_-^\mu \hat a^\dagger
+
\hat{\sigma}_+^\mu \hat a
\right)
\Bigg]
\\
&+\hat{\mathrm{H}}_{DCE}
\end{split}
\label{H largo}
\end{equation}
where
$
\hat N=\hat a^\dagger\hat a
$,
$
\Delta=\omega-\omega_d/2
$
and
$
\Delta^\mu=\omega_q^\mu-\omega_d/2
$.
The DCE contribution is given by

\begin{equation}
\hat{\mathrm{H}}_{DCE}
=
\alpha(t)
\left(
\hat a^{\dagger2}e^{i\omega_d t}
+
\hat a^2 e^{-i\omega_d t}
+
\hat a\hat a^\dagger
+
\hat a^\dagger\hat a
\right).
\end{equation}

Under the resonance condition $\omega_d=2\omega$, rapidly oscillating terms can be neglected within the rotating-wave approximation, yielding the effective Hamiltonian
\begin{equation} \begin{split} \hat{\mathcal{H}}=\Delta \hat{N} + \sum_{\mu=1}^2 \left[ \Delta^\mu \frac{\hat{\sigma}_z^\mu}{2} + g_\mu \left( \hat{\sigma}_-^\mu \hat{a}^\dagger + \hat{\sigma}_+^\mu \hat{a} \right) \right] \\ + \frac{\alpha_0}{2} \Theta(\tau-t) \left( \hat a^{\dagger 2} + \hat a^2 \right). \label{H final} \end{split} \end{equation}

Throughout the remainder of this work, the system dynamics is described by the effective Hamiltonian of Eq.~\eqref{H final}, which incorporates both the qubit -cavity interaction and the parametric driving associated with the DCE.

\subsection{Open systems description}

To account for dissipation and decoherence, we describe the dynamics within the standard Lindblad master-equation formalism \cite{linblad}. Under the usual Born, Markov, and secular approximations, the reduced density 
matrix $\hat{\rho}$ of the qubit-cavity system evolves 
according to

\begin{equation}
    \frac{d\hat{\rho}}{dt} = -i\left[ \hat{\mathcal{H}}(t), \hat{\rho}\right] 
+ \sum_{k=1}^{8} \mathcal{D}[\hat{L}_k]\hat{\rho},
    \label{eq:master_equation}
\end{equation}
where $ \hat{\mathcal{H}}(t)$ is the effective Hamiltonian in the 
rotating frame given by Eq.~(8), and the dissipator 
$\mathcal{D}[\hat{L}_k]\hat{\rho}$ for a generic collapse 
operator $\hat{L}_k$ is defined in the standard form as
\begin{equation}
    \mathcal{D}[\hat{L}_k]\hat{\rho} = \hat{L}_k \hat{\rho} 
\hat{L}_k^\dagger - \frac{1}{2} \left\{ \hat{L}_k^\dagger 
\hat{L}_k, \hat{\rho} \right\}.
\end{equation}

The set of Lindblad operators $\{\hat{L}_k\}$ models the 
primary decoherence and dissipation channels inherent to 
state-of-the-art superconducting circuit platforms. For the 
cavity resonator, we incorporate both photon loss (decay) and 
thermal excitation processes through the cavity collapse 
operators:
\begin{align}
    \hat{L}_1 &= \sqrt{\kappa_1} \hat{a}, \\
    \hat{L}_2 &= \sqrt{\kappa_2} \hat{a}^\dagger,
\end{align}
where the effective rates are given by $\kappa_1 = 
(1 + n_{\text{th}})\kappa$ and $\kappa_2 = n_{\text{th}}
\kappa$. Here, $\kappa$ is the bare cavity decay rate and 
$n_{\text{th}}$ represents the mean thermal photon number of 
the environmental reservoir at temperature $T$, determined by 
the Bose-Einstein distribution
\begin{equation}
    n_{\text{th}} = \left[ \exp\left( \frac{\hbar\omega}
{k_{\text{B}} T} \right) - 1 \right]^{-1}.
\end{equation}

For each superconducting transmon qubit ($\mu \in \{1, 2\}$), 
we consider energy relaxation, thermal excitation, and pure 
dephasing. These local decoherence mechanisms are described by 
the remaining six collapse operators:
\begin{align}
    \hat{L}_{3,\mu} &= \sqrt{\gamma_1} \hat{\sigma}_-^{(\mu)}, \\
    \hat{L}_{4,\mu} &= \sqrt{\gamma_2} \hat{\sigma}_+^{(\mu)}, \\
    \hat{L}_{5,\mu} &= \sqrt{\Gamma} \hat{\sigma}_z^{(\mu)},
\end{align}
where $\hat{\sigma}_-^{(\mu)}$, $\hat{\sigma}_+^{(\mu)}$, 
and $\hat{\sigma}_z^{(\mu)}$ are the standard lowering, 
raising, and Pauli-$Z$ operators acting on the $\mu$-th qubit. 
The qubit decay and excitation rates are parameterized as 
$\gamma_1 = (1 + n'_{\text{th}})\gamma$ and $\gamma_2 = 
n'_{\text{th}}\gamma$, respectively, with $\gamma$ being the 
bare relaxation rate and $n'_{\text{th}}$ denoting the thermal 
population of the qubit's local bath. Collectively, these 
formulations define the eight collapse operators acting on the 
composite Hilbert space of our system, enabling a comprehensive 
numerical study of the resulting synchronization dynamics under 
realistic environmental conditions.

In the next section, we analyze the resulting synchronization dynamics by numerically solving Eq. (\ref{eq:master_equation}) under different environmental conditions.

\section{DCE-INDUCED SYNCHRONIZATION}

In this section, we investigate whether the dynamical Casimir effect can induce synchronization between two initially uncorrelated qubits and analyze the robustness of this synchronization against environmental effects.
We first consider the closed-system dynamics, which provides a reference for the dissipative cases considered below.

Since synchronization in quantum systems cannot be characterized by a unique observable, several measures based on the dynamics of local observables and quantum states have been proposed \cite{syncdinamic1,syncdinamic2,syncdinamic3,luciernagas,sincro1,sincro2,sincro3}. Among them, the Pearson correlation coefficient is one of the most widely used indicators of synchronized dynamics \cite{pearson}. In the present work, synchronization is characterized by the Pearson coefficient evaluated from the local qubit observables $\langle \hat{\sigma}_z^{(1)}\rangle$ and $\langle \hat{\sigma}_z^{(2)}\rangle$, defined as

\begin{equation}
C_{\Delta t}(t)=
\frac{
\int_t^{t+\Delta t}
\delta \sigma_z^{(1)}(t')\,\delta \sigma_z^{(2)}(t')\,dt'
}{
\sqrt{
\int_t^{t+\Delta t}
\left[\delta \sigma_z^{(1)}(t')\right]^2dt'
\int_t^{t+\Delta t}
\left[\delta \sigma_z^{(2)}(t')\right]^2dt'
}
},
\label{pearson}
\end{equation}

\noindent where $
\delta \sigma_z^{(i)}(t)=
\langle \hat{\sigma}_z^{(i)}(t)\rangle-
\overline{\sigma}_z^{(i)}$
and
 $\overline{\sigma}_z^{(i)}$ denotes the average expectation value of $\langle \hat{\sigma}_z^{(i)}\rangle$ within the time window $\Delta t$. 
 The Pearson coefficient takes values in the interval $[-1,1]$, where $+1$ ($-1$) corresponds to perfectly correlated (anticorrelated) dynamics.
  Throughout this work, the Pearson coefficient is evaluated using a time window of $\omega \Delta t=200$, following Ref.~\cite{tesisjapon}.

The system is initially prepared in the product state
\begin{eqnarray}
|\psi(0)\rangle &=&
(\cos\theta_1 |g\rangle_1+\sin\theta_1 |e\rangle_1) \\
 &\otimes &
(\cos\theta_2 |g\rangle_2+\sin\theta_2 |e\rangle_2)
\otimes |0\rangle \nonumber,
\end{eqnarray}
where $|g\rangle$ and $|e\rangle$ denote the ground and excited states of each qubit, respectively, and $|0\rangle$ represents the cavity vacuum state.

Unless otherwise stated, all frequency parameters and coupling strengths throughout this work are expressed in units of the cavity fundamental frequency $\omega$ (setting $\omega=1$) rendering them dimensionless. Thus, the resonant qubit frequencies are
$\omega=\omega_q=1$ and the coupling constants are
$g_1=g_2=0.04 \omega$. Initial qubit states are chosen with $\theta_1=\pi/4$ and $\theta_2=11\pi/36$. The DCE modulation amplitude is fixed at $\alpha_0=6\times10^{-3} \omega$ (Ref.~\cite{tesisjapon}). The parametric modulation is switched off at $\omega \tau=500$. This choice is motivated by the continuous generation of photons inside the cavity by the DCE, which leads to a rapid increase in the Hilbert-space dimension required for accurate numerical simulations. Limiting the pumping duration therefore allows us to access the long-time dynamics while keeping the computational cost under control.

\subsection{Role of dephasing and dissipation}

\begin{figure}
\centering
\includegraphics[width=1.0\linewidth]{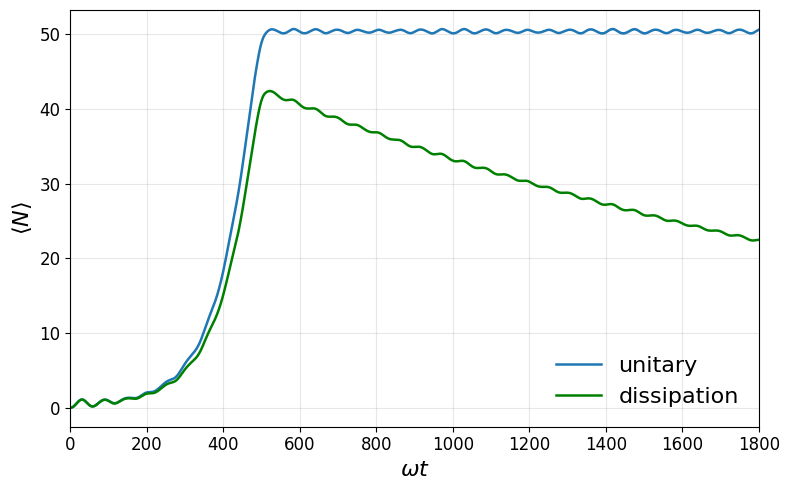}
\caption{Mean cavity photon number as a function of time. The blue curve corresponds to the unitary dynamics under the resonance condition $\omega_d=2\omega$, while the green curve includes dissipation. It is easy to note that the inclusion of environmental losses suppresses the photon population generated by the DCE. Parameters used: $\omega=\omega_q=1$, $
g_1=g_2=0.04 \omega$, $\theta_1=\pi/4$ and $\theta_2=11\pi/36$ for the systemn, and  $\kappa=0.0125g$ for the environmental losses.}
\label{numero}
\end{figure}

\begin{figure*}[t]
\centering

\begin{subfigure}{0.48\textwidth}
\centering
\includegraphics[width=\textwidth]{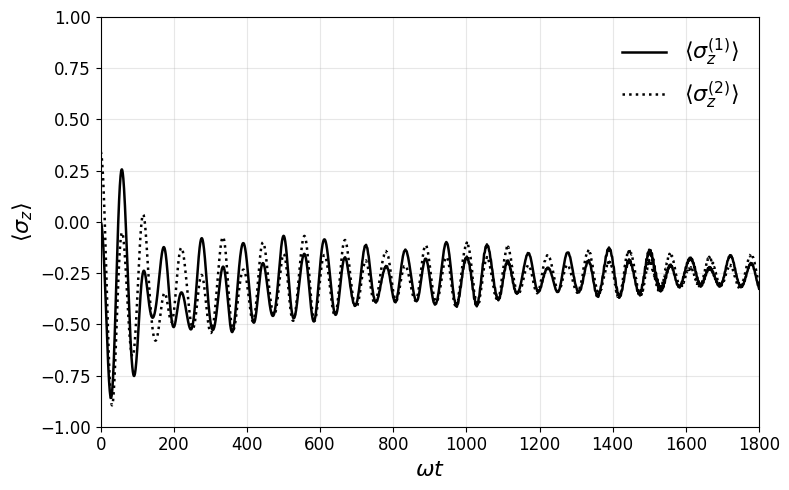}
\caption{Expectation value of $\sigma_z$ as a function of time in the presence of pure dephasing. The oscillation amplitude gradually decreases due to the loss of quantum coherence induced by the dephasing process. The dephasing rate is set to $\Gamma = 0.0125g$.}
\label{sigmadep}
\end{subfigure}
\hfill
\begin{subfigure}{0.48\textwidth}
\centering
\includegraphics[width=\textwidth]{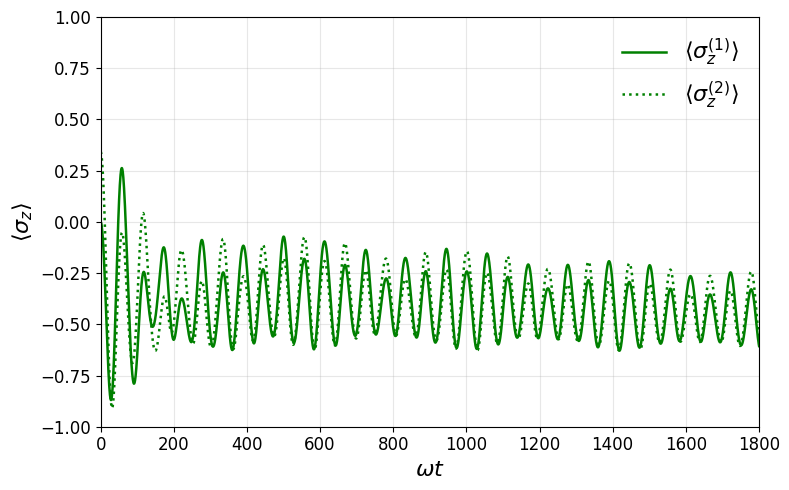}
\caption{Expectation value of $\sigma_z$ as a function of time in the presence of cavity losses. Photon dissipation suppresses the cavity-mediated dynamics and drives the qubits toward their ground state. The dissipation rate is set to $\kappa = 0.0125g$.}
\label{sigmadip}
\end{subfigure}
\caption{Time evolution of the expectation value $\langle \sigma_z \rangle$ under different environmental decoherence mechanisms. Panel (a) shows the effect of pure dephasing, while panel (b) illustrates the effect of cavity photon losses.}
\label{sigmaz}
\end{figure*}

We independently analyze the effects of cavity photon losses and pure dephasing on DCE-induced synchronization. This comparison allows us to separately assess the impact of photon losses and loss of quantum coherence on synchronization. The dynamics are described by the cavity-loss and pure-dephasing channels of the Lindblad master equation, with rates $\kappa=0.0125g$ and $\Gamma=0.0125g$, respectively. Throughout this section, the temperature is set to zero in order to isolate the effects of each decoherence mechanism.

For the dissipative case, we first plot the average photon number as a function of time and compare it with the unitary dynamics. Fig. \ref{numero} shows this quantity for both cases, with the unitary dynamics represented in blue and the dissipative dynamics in green. In the unitary case, the photon population grows during the pumping stage as a consequence of the continuous generation of photon pairs by the DCE until $\omega t=500$, when the pumping is switched off. After this point, the number of photons remains approximately constant, reaching about 50 photons inside the cavity. In the dissipative case, the photon number also grows during the pumping phase, albeit reaching a lower peak at $\omega t = 500$ than in the unitary evolution. Once the DCE is turned off, no additional photons are generated, and the photons already present in the cavity begin to dissipate. This reduction in the intracavity photon population directly weakens the cavity-mediated interaction responsible for transferring correlations between the qubits.

\begin{figure}
\centering
\includegraphics[width=1\linewidth]{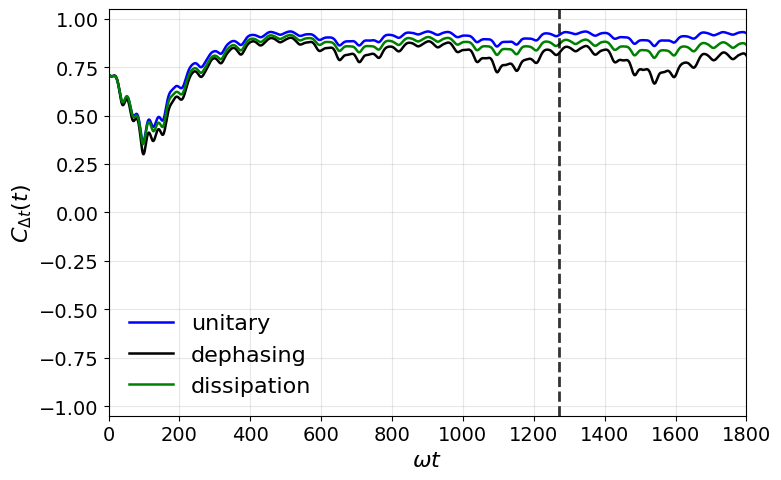}
\caption{Pearson correlation coefficient $C_{\Delta t}$ as a function of time for different dissipation mechanisms. The blue curve corresponds to the unitary evolution, the green curve to the case with cavity photon losses, and the black curve to the pure dephasing regime.  The vertical dashed lines indicate the corresponding decoherence times whenever they lie within the simulated time window. The simulation parameters are the same as those used in Fig.~\ref{sigmaz}.}
\label{pearson_aislado}
\end{figure}

In Fig. \ref{sigmaz}, we show the expectation values of $\sigma_z$ for the dissipative and pure-dephasing cases, respectively, as a function of time. Unlike in Fig.~\ref{numero}, the unitary evolution is not shown here. In the dissipative case, the expectation values of $\sigma_z$ tend to decay over time. This behavior is associated with photon losses in the system, which eventually drive the qubits towards their ground states. In contrast, under pure dephasing, the expectation values remain centered around the same equilibrium value, while their oscillations gradually lose amplitude because the environment suppresses the phase coherence between the qubit states, even though no energy is exchanged with the environment. The impact of the environment on synchronization is quantified through the Pearson correlation coefficient shown in Fig.~\ref{pearson_aislado}.

Fig. \ref{pearson_aislado} shows the Pearson correlation coefficient for the unitary, dissipative, and pure-dephasing cases. The vertical black line indicates the characteristic decoherence time for the pure-dephasing dynamics. To quantify the loss of quantum coherence, we define the decoherence time from the decay of the off-diagonal elements of the reduced two-qubit density matrix. Specifically, we track several coherences $|\rho_{ij}(t)|$ and define $t_d$ decoherence time as the instant at which

\begin{equation}
|\rho_{ij}(t_d)|=
\frac{|\rho_{ij}(0)|}{e}.
\label{decoherence_time}
\end{equation}

Throughout this work, $t_d$ is defined as the largest decay time to $1/e$ of the initial value among the selected coherences, thus providing a conservative estimate of the persistence of quantum coherence.

The blue curve, corresponding to the unitary dynamics, attains the highest values. The correlation increases during the first 500 units of time and subsequently oscillates around an approximately constant value. The green curve corresponds to the dissipative case and follows a behavior similar to the unitary dynamics, although with slightly lower correlation values. Interestingly, cavity losses only weakly affect the synchronization dynamics. Although photons are continuously removed from the cavity, the Pearson coefficient remains close to the unitary value over long times, indicating that the correlations associated with synchronization are relatively robust against moderate dissipation. Finally, the black curve corresponds to the pure-dephasing case. A marked reduction in the Pearson coefficient is observed after the decoherence time, indicating that synchronization rapidly degrades once the longest-lived quantum coherences have decayed. These results suggest that both dissipation and dephasing reduce synchronization between the qubits, with dephasing being the dominant mechanism responsible for the loss of correlation.

\subsection{Synchronization in realistic cQED platforms}

In this section, we investigate whether synchronization induced by the dynamical Casimir effect can be observed in current circuit quantum electrodynamics (cQED) platforms. In all cases, we consider a temperature of $60 \mathrm{mK}$, representative of typical experimental conditions. Our goal is to determine to what extent the losses and decoherence mechanisms present in realistic devices affect DCE-induced synchronization and the quantum correlations between the qubits.

To this end, we consider three representative dissipation regimes, which we refer to as strong, intermediate, and weak. In the strong regime, we set $\kappa=\gamma=0.01 g$ and $\Gamma=0.1g$. In the intermediate regime, we set $\kappa=\gamma=0.000.5 g$ and $\Gamma=0.01g$, while in the weak regime we use $\kappa=\gamma=0.0002g$ and $\Gamma=0.004g$. The first two parameter sets were chosen based on experimental values reported for conventional cQED devices \cite{tasafuerte1,tasafuerte2,tasafuerte3,tasafuerte4}, while the weak regime corresponds to dissipation rates of the order of those achieved in state-of-the-art experimental implementations \cite{nicoferpaz,tasadebil1,tasadebil2}. In this way, the three regimes cover a representative range of experimental parameters, from conventional cQED implementations to devices with the lowest dissipation rates reported to date. In all cases, the remaining model parameters are kept fixed in order to isolate the effect of the different dissipation rates.

\begin{figure}
\centering
\includegraphics[width=1\linewidth]{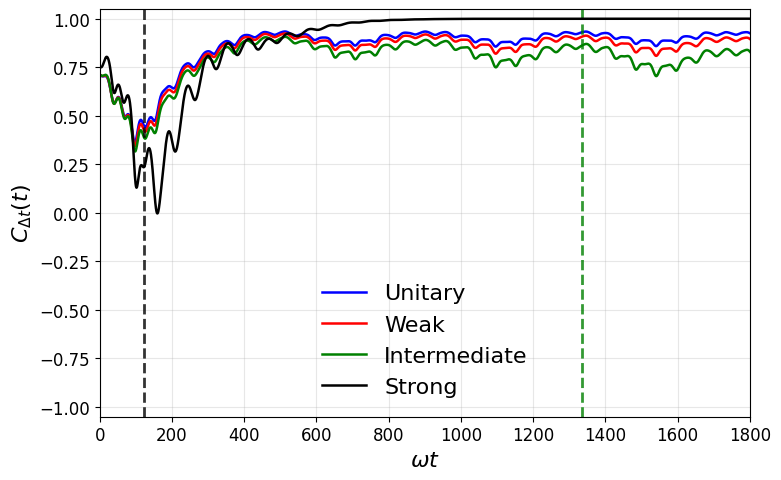}
\caption{Pearson correlation coefficient as a function of time for the four dynamical regimes considered. The blue curve corresponds to the unitary evolution, while the red, green, and black curves represent the weak-, intermediate-, and strong-dissipation regimes, respectively. The vertical dashed lines indicate the corresponding decoherence times whenever they lie within the simulated time window. The parameters are $\kappa=\gamma=0.0002g$ and $\Gamma=0.004g$ for the weak regime, $\kappa=\gamma=0.0005g$ and $\Gamma=0.01g$ for the intermediate regime, and $\kappa=\gamma=0.01g$ and $\Gamma=0.1g$ for the strong regime.}
\label{pearson_new}
\end{figure}

Fig. \ref{pearson_new} shows that the weak-dissipation regime is almost indistinguishable from the unitary evolution, demonstrating that DCE-induced synchronization is robust against dissipation levels currently achieved in state-of-the-art cQED devices. In this figure, we compare the time evolution of the Pearson correlation coefficient for the four dynamical regimes considered: the unitary evolution (blue curve) and the weak (red), intermediate (green), and strong (black) dissipation cases. The vertical dashed lines indicate the corresponding decoherence times whenever they fall within the simulated time window.

For the unitary dynamics and the weak-dissipation regime, the Pearson coefficient exhibits almost identical behavior. After an initial transient, the Pearson coefficient oscillates around a value close to unity, indicating the establishment of a synchronized regime. The small differences between these two curves show that the synchronization mechanism is remarkably robust against the weak dissipation considered here. The decoherence time $t_d$ is not shown because it exceeds the time window considered.

As the environmental coupling increases, the behavior changes qualitatively. In the intermediate regime, the Pearson coefficient develops larger oscillations and its average value is reduced, indicating that synchronization progressively deteriorates due to decoherence. Nevertheless, significant positive correlations remain even after the decoherence time.

In contrast, the strong-dissipation regime shows a Pearson coefficient that remains close to unity over most of the evolution, apparently indicating stronger synchronization than in the unitary case. Since such behavior is counterintuitive, the Pearson coefficient alone cannot determine whether genuine quantum synchronization persists. This issue is addressed in the next subsection through an analysis of quantum discord.

\subsection{Quantum correlations beyond Pearson synchronization}

To clarify the unexpectedly large Pearson coefficient in the strong-dissipation regime, we complement the synchronization analysis with two standard quantifiers of quantum correlations: Concurrence and Quantum discord. Although these quantities do not measure synchronization itself, they characterize the nonclassical correlations between the qubits and have been widely employed to analyze quantum synchronization \cite{sincro1,sincro2}.

For a two-qubit system, the Concurrence is defined as \cite{calcular_concurrence}
\begin{equation}
\mathcal{C}(\hat{\rho})
=
\max\left\{
0,\lambda_1-\lambda_2-\lambda_3-\lambda_4
\right\},
\label{concurrence}
\end{equation}
where $\lambda_i$ are the square roots of the eigenvalues of
$\hat{\rho}\hat{\rho}_{\mathrm{sf}}$, arranged in decreasing order, with
\[
\hat{\rho}_{\mathrm{sf}}
=
(\hat{\sigma}_y\otimes\hat{\sigma}_y)
\hat{\rho}^{*}
(\hat{\sigma}_y\otimes\hat{\sigma}_y).
\] The Concurrence ranges from $0$ for separable states to $1$ for maximally entangled states.

Quantum discord captures more general forms of quantum correlations, including those present in separable states, and has been shown to play an important role in quantum information processing \cite{discord1,discord2}. It is defined as the difference between the total correlations $I(\hat{\rho}_{ab})$ and the classical correlations $J_{a:b}(\hat{\rho})$ \cite{discord3,discord4}. The quantum discord is therefore given by
\begin{equation} \delta_{a:b}(\hat{\rho}) = I(\hat{\rho}_{ab}) - J_{a:b}(\hat{\rho}). \end{equation}
This quantity can be interpreted as the minimum discrepancy between the two quantum extensions of classical mutual information, obtained by optimizing over the set of measurements \cite{sincro2}.

\begin{figure}
\centering
\includegraphics[width=1.0\linewidth]{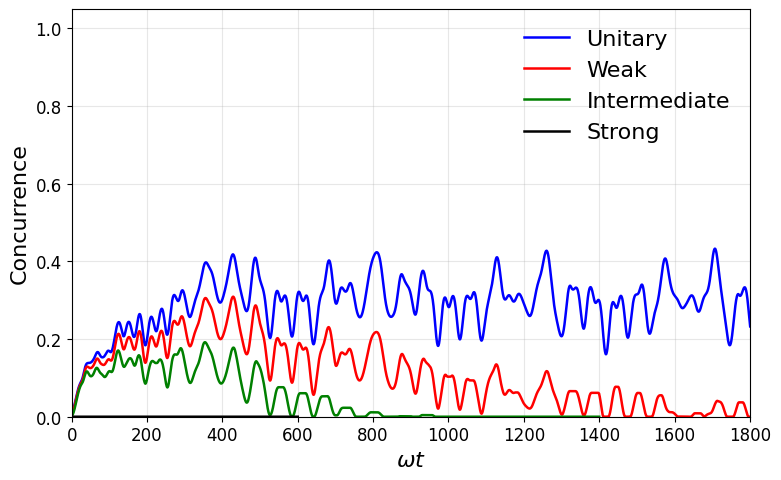}
\caption{Time evolution of the Concurrence for the unitary evolution and the three dissipative regimes. For visualization purposes, the data were smoothed using a moving average over 60 consecutive points. The parameters are chosen as in Fig.~\ref{pearson_new}.}
\label{concurrence_new}
\end{figure}

Fig. \ref{concurrence_new} compares the Concurrence for the four dynamical regimes considered in the previous section. In the weak-dissipation regime, the concurrence remains close to its unitary value throughout the simulated time interval, showing that entanglement is only weakly affected by the environment. As the dissipation increases, entanglement becomes progressively more fragile. In the intermediate regime, the concurrence rapidly decreases and vanishes shortly before the decoherence time, whereas in the strong-dissipation regime it is almost completely suppressed at very early times. These results show that entanglement is highly sensitive to environmental losses and that the temporal window over which it survives becomes increasingly limited as the dissipation strength increases.

\begin{figure}
\centering
\includegraphics[width=1.0\linewidth]{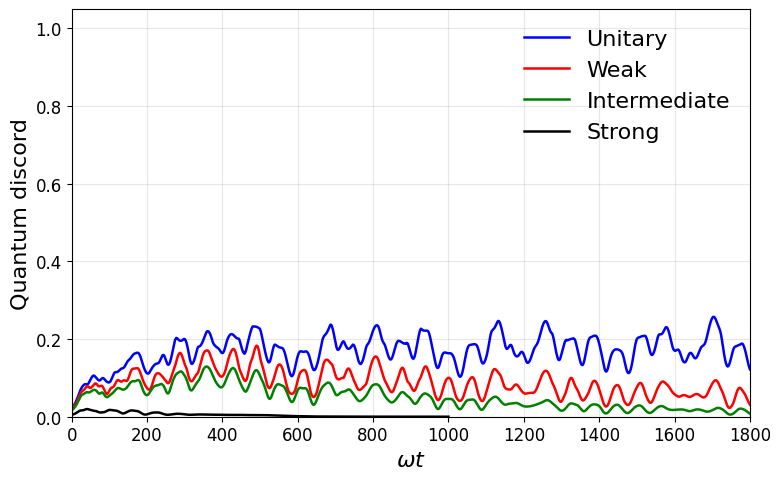}
\caption{Time evolution of the Quantum discord for the unitary evolution and the three dissipative regimes. For visualization purposes, the data were smoothed using a moving average over 60 consecutive points.  The parameters are chosen as in Fig.~\ref{pearson_new}.}
\label{discord_new}
\end{figure}

The corresponding quantum discord is shown in Fig.~\ref{discord_new}. As dissipation increases, quantum correlations are progressively reduced. However, quantum discord exhibits significantly greater robustness against decoherence. While concurrence disappears in the intermediate regime, quantum discord remains finite throughout the entire evolution, indicating that nonclassical correlations persist even after entanglement has been completely destroyed. Only in the strong-dissipation regime does quantum discord rapidly decay to negligible values. The greater robustness of quantum discord against environmental effects is consistent with previous studies \cite{sincro2}.

Comparing Figs.~\ref{pearson_new}, \ref{concurrence_new}, and \ref{discord_new} allows us to identify the physical origin of the Pearson correlations. In the weak-dissipation regime, the large Pearson coefficient is accompanied by significant quantum correlations, indicating genuine DCE-induced quantum synchronization. In the intermediate regime, synchronization survives for a finite time while the quantum correlations become progressively weaker. Finally, in the strong-dissipation regime, the persistence of a Pearson coefficient close to unity despite the almost complete disappearance of both concurrence and quantum discord demonstrates that the observed synchronization is not of quantum origin. Instead, it arises from the dissipative relaxation of both qubits toward the same stationary state.

These results demonstrate that a large Pearson coefficient does not necessarily imply genuine quantum synchronization. In the strong-dissipation regime, the observed correlations originate from the common dissipative relaxation of both qubits rather than from DCE-induced quantum correlations. Therefore, the Pearson-based synchronization analysis should be complemented by quantum-correlation measures when environmental dissipation becomes significant.

\section{Conclusions}

In this work, we investigated synchronization induced by the dynamical Casimir effect in a circuit quantum electrodynamics (cQED) architecture consisting of two superconducting qubits coupled to a parametrically driven cavity. By combining a Lindblad master-equation approach with different measures of synchronization and quantum correlations, we analyzed how environmental decoherence affects the emergence and persistence of quantum synchronization.

We first examined separately the effects of cavity photon losses and pure dephasing. While both mechanisms reduce synchronization, our results show that pure dephasing has a considerably stronger effect on the Pearson correlation coefficient, indicating that the loss of quantum coherence is the dominant mechanism responsible for the degradation of synchronization.

We then analyzed three experimentally relevant dissipation regimes representative of current cQED platforms. For weak dissipation, DCE-induced synchronization remains essentially unaffected and coexists with significant quantum correlations, suggesting that this phenomenon should be experimentally observable in state-of-the-art superconducting circuits. As the dissipation increases, quantum correlations become progressively more fragile than synchronization, leading to an intermediate regime in which synchronization persists only within a finite temporal window limited by decoherence. Finally, for sufficiently strong dissipation, the Pearson correlation coefficient remains close to unity even after both concurrence and quantum discord have nearly vanished. This demonstrates that, in this regime, the observed correlations originate from environment-induced relaxation toward a common stationary state rather than from genuine quantum synchronization.

These results highlight an important practical implication for experiments. Although the Pearson correlation coefficient is a useful indicator of synchronized dynamics, it is not by itself sufficient to identify synchronization of quantum origin. In experimentally relevant parameter regimes, it should therefore be complemented by measures of quantum correlations, such as concurrence and quantum discord, in order to correctly identify the physical mechanism responsible for the observed correlations.

Overall, our results provide a systematic characterization of DCE-induced synchronization under realistic experimental conditions and establish the parameter regimes in which genuine quantum synchronization can be reliably identified in superconducting cQED platforms. We hope that these findings contribute to future experimental studies of quantum synchronization and to the development of quantum technologies based on dynamically generated quantum correlations.

\section*{Acknowledgments}
This work was supported by Consejo Nacional de Investigaciones
Científicas y Técnicas (CONICET) and Universidad de
Buenos Aires (UBA).

\newpage

\bibliography{references}

\end{document}